\documentclass[prl,aps]{revtex4}
\usepackage{amsmath}
\usepackage{amsfonts}
\usepackage{amssymb}
\usepackage{graphicx}
\usepackage{color}

\begin{document}

\title{Long-range spin correlations in a  honeycomb spin model with magnetic field}
\author{A. V. Lunkin$^{1,2}$, K. S. Tikhonov$^{1,2}$ and  M. V. Feigel'man$^{1,2}$ }
\affiliation{$^1$ L. D. Landau Institute for Theoretical Physics, Kosygin str.2, Moscow
119334, Russia}
\affiliation{$^2$ Moscow Institute of Physics and Technology, Moscow 141700, Russia}
\date{\today }

\begin{abstract}
We consider spin-$\frac{1}{2}$ model on the honeycomb lattice~\cite{Kitaev06}
in the presence of weak magnetic field $h\ll J$. 
Such a perturbation treated in the second order over $h$
leads~\cite{TFK2011} to the power-law decay of irreducible spin correlation function
 $S(\mathbf{r},t)=\left\langle \left\langle s^{z}_r(t)s^{z}_0(0)\right\rangle \right\rangle
\propto h_{z}^{2}f(t,\mathbf{r})$, where $f(t,\mathbf{r})\propto \lbrack \max (t,Jr)]^{-4}$ 
is an oscillating function of $\mathbf{r}$,
with a wavelength equal to 3 lattice constants.
In the present Letter we sum main terms in all orders of the perturbation theory for the correlation function 
$S(\mathbf{r},t)$ in the limit of large $r,t$.  Our results can be understood 
in terms of the effective low-energy Hamiltonian written in terms of Majorana fermions, which in the presence of magnetic field acquire vector potential $A_x \propto  h_z^2$. Correspondingly, the wave vector of the oscillations in
 $S(\mathbf{r},t)$ changes according to  $\delta k \propto h_z^2$.
 We also compute the dynamic structure factor $S(\mathbf{p},\omega)$;  in the vicinity of $\mathbf{p}_K$ corresponding to the inter-conical points excitations it reads as $S(\mathbf{p},\omega)-S(\mathbf{p}_K,\omega)\propto\sqrt{\omega^2-3J^2(\mathbf{p}-\mathbf{p}_K)^2}$.

\end{abstract}

\maketitle

\emph{Introduction}

Quantum spin liquids, QSL's (see e.g. Refs.~\cite{PWA1,PWA2,Wen,rev1,rev2}) present examples of strongly correlated quantum phases which do not develop any kind of local order, while their specific entropy vanishes at zero
temperature. \emph{Critical}, or algebraic QSL's are characterized by spin
correlation functions that decay as some power of distance and time.
One exactly solvable case of the critical QSL is presented by the celebrated Kitaev honeycomb spin model~\cite{Kitaev06}, for a more recent review see Ref. \onlinecite{nussinov2013}. Although long-range spin correlations exactly vanish in this model, it presents convenient starting point for the construction of controllable theories possessing long-range spin correlations,
since the spectrum of the model contains gap-less fermions. 
Honeycomb model~\cite{Kitaev06}  was originally invented as a simplest solvable spin model
possessing nontrivial topological phases, relevant in the context of topological quantum computing; 
later it has been  found that similar spin interactions
can be realized in the honeycomb-lattice oxides Na$_2$IrO$_3$ and Li$_2$IrO$_3$\cite{jackeli09,brink14,katukiri15}. 
In a realistic situation, low-energy effective description of these materials is given by a mixture of the Kitaev and Heisenberg interactions with weights depending on the microscopic parameters. 
Alternatively, Heisenberg-Kitaev (HK) model appears as a low-energy theory of a Hubbard model defined on a honeycomb lattice with spin-dependent hopping\cite{hassan13}. Interestingly enough,
exact diagonalization and a complementary spin-wave analysis\cite{chaloupka10} show that spin-liquid phase near the Kitaev limit is stable with respect to small admixture of Heisenberg interactions\cite{shaffer12}. 

However Kitaev model in its original form does not possesses long-range spin correlations, moreover, its spin correlators 
are strictly local\cite{Baskaran07}. A perturbative  addition of the  Heisenberg interaction  does not change this fact
\cite{baskaran11}.  In order to produce a spin-liquid phase  with long-range correlations, some other terms should be added to the effective Hamiltonian.  In particular, such a terms appear naturally if HK Hamiltonian is obtained as 
a low-energy limit for the Hubbard model\cite{hassan13}.  Another perturbation which does not destroy spin-liquid phase but renders correlations non-local is magnetic field\cite{TFK2011, horsch11}.

Importantly, the case of a weak magnetic field $h_z\sigma_i^z$  added to the Kitaev model is tractable analytically and in the paper~\cite{TFK2011} it was shown that indeed algebraic QSL can be obtained as a result of such a simple perturbation applied to the Kitaev model. It was found\cite{TFK2011} that it leads to an appearance of
long-range (power-law) contribution to the irreducible spin-spin correlation function $
S(r,t)=\left\langle\left\langle s_{r}^{z}(t)s_{0}^{z}(0)\right\rangle\right\rangle,$
where  $s_r = \sigma_{r,1} + \sigma_{r,2}$ is the total spin of an elementary cell.
This result was obtained in the leading non-vanishing order in the perturbation strength: power-law contribution to $S$ is proportional to $h_z^2$.
Qualitatively,  the result of Ref.~\cite{TFK2011} can be interpreted in very simple terms:
magnetic field provides a coupling between the spin operator and the operator of  density of 
Majorana fermions which are used to diagonalize the unperturbed ($h_z=0$) Hamiltonian (see below). Once this coupling was demonstrated, the rest of the calculation is rather
straightforward: one should calculate density-density correlation function for these free
fermions. 

However, it worked this way in the lowest order in $h_z$ only.  Should one be interested in the effects
of higher order in the magnetic field,  its influence upon the properties of Majorana fermions should be 
studied. This is the subject of the  present Letter:  we demonstrate how to calculate spin-spin correlation
function in the next ($h_z^4$) order and then show that the result can be understood in a very simple
terms of rather natural perturbation applied to the free Majorana problem.

We consider the model defined by the Hamiltonian: 
\begin{equation}
\mathcal{H}=J\sum_{l=\left\langle ij\right\rangle }\left( \mathbf{\sigma }%
_{i}\mathbf{n}_{l}\right) \left( \mathbf{\sigma }_{j}\mathbf{n}_{l}\right)
-\sum_{i}\mathbf{h}_{i}\mathbf{\sigma }_{i}.  \label{H}
\end{equation}%
Unit vectors $\mathbf{n}_{l}$ are parallel to $x$, $y$ and $z$ axis for the
corresponding links $x$, $y$ and $z$ of the honeycomb lattice:
$\mathbf r=m_1 \mathbf n_1 + m_2 \mathbf n_2$ with integer $m_{1,2}$ and translation vectors 
$n_{1,2}=(\pm\frac{1}{2},\frac{\sqrt3}{2})$.
 At
$\mathbf{h}_{i}\equiv 0$ the Hamiltonian (\ref{H}) was solved
exactly~\cite{Kitaev06} via a mapping to a free fermion Hamiltonian. In this
approach, each spin $\sigma _{i}$ is represented in terms of four Majorana
operators $c_{i},~c_{i}^{x},~c_{i}^{y},~c_{i}^{z}$ with the following
anticommutation relations: $\bigl\{ c_{i}^{\alpha },c_{j}^{\beta }\bigr\}
=2\delta _{ij}\delta _{\alpha \beta }$, so that $\sigma _{i}^{\alpha
}=ic_{i}c_{i}^{\alpha }$. In terms of these new operators, the zero-field
Hamiltonian reads $\mathcal{H}=-iJ\sum_{\left\langle ij\right\rangle
}c_{i}u_{ij}c_{j}$ $\ $and $u_{ij}=ic_{i}^{\alpha }c_{j}^{\alpha }$ are
constants of motion: $\left[ \mathcal{H},u_{ij}\right] =0$, with $u_{ij}=\pm 1
$. The ground state $|G\rangle$ corresponds to a choice of
$\left\{u_{ij}\right\}$ that minimizes the fermionic energy. It is convenient to
introduce the notion of $Z_{2}$ flux, defined for each hexagon $\pi$ as a
product $\phi_{\pi}=\prod u_{ij}$ (since $u_{ij}=-u_{ji}$, we have to choose a
particular ordering in this definition: $i\in\text{even sublattice}$,\,
$j\in\text{odd sublattice}$).  The ground state of this model is a symmetrized
sum of states with different sets of integrals of motion $\left\{
  u_{ij}\right\}$, corresponding to all fluxes equal to $1$. Such a symmetrization, however, never needs to be implemented in practice and $S(r, t)$ can be 
computed with unprojected eigenstates. This is possible due to gauge invariance of the spin operators
(in general, one should take care of the parity of fermions in the physical sector, which can depend on the boundary conditions\cite{loss11}).

Fixing the gauge (all $u_{ij}\equiv 1$), we denote by $H$ the
corresponding Majorana Hamiltonian: $H=-iJ\sum_{\left\langle ij\right\rangle
}c_{i}c_{j}$. It can be diagonalized with the use of Fourier
transformation; as a result  the spectrum of unperturbed Kitaev model reads $\epsilon(p)=\pm|f(\mathbf p)|$ where
 $f(\mathbf p)=2iJ(1+e^{i(\mathbf p,\mathbf n_1)}+e^{i(\mathbf p,\mathbf n_2)})$. 
It is gapless and has two conic points (lattice constant is set to be unity):
\begin{equation}
\mathbf{K}_{1,2}=(\pm\frac{2\pi}{3},\frac{2\pi}{\sqrt{3}})
\label{Kpoints}
\end{equation}

Long-range correlation functions are determined by the fermionic fields with momenta close to
either $ \mathbf{K}_1$ or $\mathbf{K}_2 $. Our original site fermions $c_{i\lambda}$ are real
(here $i$ enumerates elementary cells while $\lambda=1,2$ selects one of the two  sublattices),
and after Fourier transform we have $c^+_\lambda(\mathbf{q}) = c_\lambda(-\mathbf{q})$.
As long as we are interested in low-energy behaviour, it is possible to work with complex
fermionic fields $a_{\lambda}(\mathbf{p})$ and $a^+_{\lambda}(\mathbf{p})$ with small
 momentum $\mathbf{p}$ defined as follows:
\begin{eqnarray}
\label{ac}
a_{\lambda}(\mathbf{p}) =  c_\lambda(\mathbf{K}_1 + \mathbf{p}) \\ \nonumber
a^+_{\lambda}(\mathbf{p}) = c_\lambda(\mathbf{K}_2 - \mathbf{p})
\end{eqnarray}
In terms of $a_\lambda(\mathbf{p})$ fermions  our problem can be formulated in a continuous form, without
 reference to the underlying lattice.
Now we  introduce $2\times 2$ Pauli  matrices $\sigma^\alpha_{\lambda\mu}$, acting in the sublattice space,
with $\alpha = x$ or $y$,
and present our  low-energy Hamiltonian in the Dirac form
\begin{equation}
H_F = \sqrt{3} J \sum_{\mathbf{p}} \bar{a}_\lambda(\mathbf{p})\sigma^\alpha_{\lambda\mu} p^\alpha  a_\mu(\mathbf{p})
\label{Dirac}
\end{equation}
where new "Dirac-conjugated" fields $\bar{a}_\lambda(\mathbf{p}) =
 -i a^+_\nu(\mathbf{p}) \sigma^z_{\nu\lambda}$ are introduced for convenience
 (matrix $-i\sigma^z$ is acting like charge conjugation operator).

\begin{figure}[tbp]
\includegraphics[width=5.5cm,height=5cm]{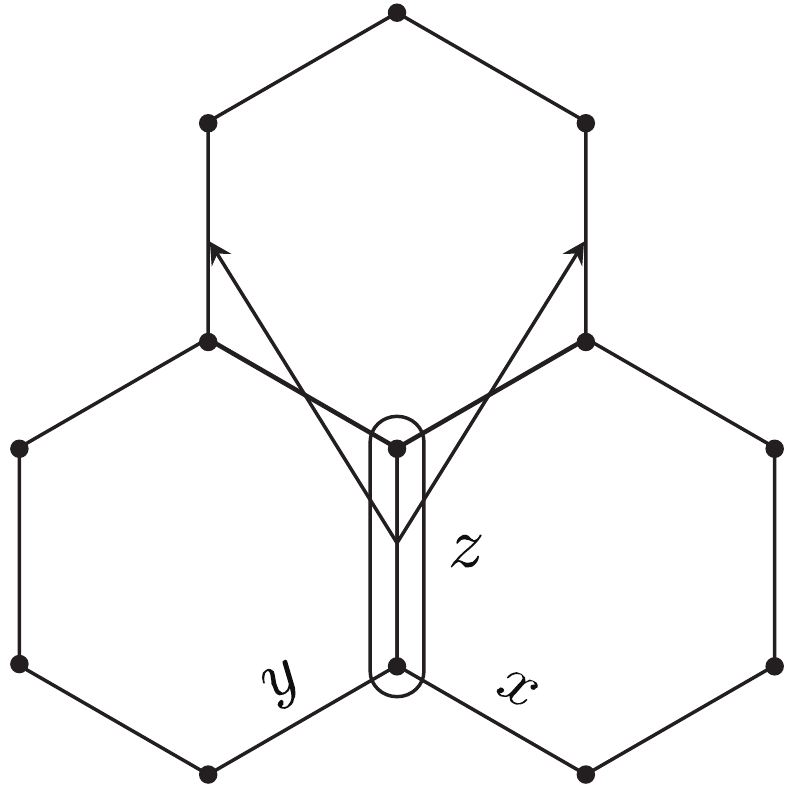}
\caption{A honeycomb lattice fragment: bond fermions $\psi, \phi$ belong to the z-link indicated on the figure.}
\label{pcell}
\end{figure}
We have shown in the previous paper~\cite{TFK2011}, that magnetic field induces the coupling of spin to Fermionic density, leading to the following result: 
\begin{equation}
\label{res2}
S^{(2)}(\mathbf{r},t) = \frac{4}{\pi^2}\left(\frac{h_z}{h_0}\right)^2\frac{3J^{2}t^{2}\left( Z_{1}^{2}+Z_{2}^{2}\right) -r ^{2}\left( Z_{1}^{2}+Z_{2}^{2}-2Z_{1}Z_{2}\cos 2\gamma \right) }{\left( 3J^{2}t^{2}-r ^{2}\right) ^{3}},
\end{equation}
where  $h_0 \sim J$ and we have introduced $Z_{1,2}=e^{i \mathbf{K}_{1,2} \mathbf{r}}$, where conical points wavevectors $\mathbf{K}_{1,2}$ are defined in Eq.(\ref{Kpoints})
 and $\gamma$ stays for the polar angle of $\mathbf r$ in the $(x,y)$.
Equivalent form of (\ref{res2}) is
\begin{equation}
S^{(2)}(\mathbf{r},t)=\frac{8}{\pi^2}\left(\frac{h_z}{h_0}\right)^2\left(\frac{3J^2t^2+r^2\cos2\gamma}{(3J^2t^2-r^2)^3}+\frac{\cos2\pi m/3}{(3J^2t^2-r^2)^2}\right)
\end{equation}
with $m = m_1-m_2$. 

Below we will employ the same  method of calculation, used in Ref.~\cite{TFK2011}
to obtain Eq.(\ref{res2}), generalizing it to a next order in  magnetic field.  
We will be interested in long-time asymptotics  of the spin-spin correlation function,  hence rich and important physics of Fermi-edge-like singularity for Majorana Fermions~\cite{moessner14,moessner15}  will be of limited importance for us;
these time-dependents effects will only restrict the relevant domain of the integration over the intermediate states in
 the corresponding perturbation theory, as described below.

\emph{Reduction of the spin-spin correlation function to fermionic one}\\
We start from the expression for the spin-spin correlation function 
$S(\mathbf{r},t)$, expanded up to the
fourth order in magnetic field $h_z$. 
It reads (compare with Eq.(2) in Ref.~\cite{TFK2011}) as follows:
\begin{equation}
\label{h4}
S^{(4)}(\mathbf{r},t) =
\frac{h_z^4}{4!}\sum\limits_{r_1\ldots r_4}\int d\tau_1 \ldots  d\tau_4 \langle T s^z_r(t)s^z_0(0)s^z_{r_1}(\tau_1)\ldots s^z_{r_4}(\tau_4)\rangle.
\end{equation}
We are interested in irreducible correlation function, so that we will have in mind that only irreducible diagrams should be taken into account.

It is convenient to introduce complex "bond fermions",
defined on $z$-links as follows:
$\psi_r = \frac12\left(c_{r1} + ic_{r2}\right)$ and $\phi_r=\frac{1}{2}(c_{r,1}^z+ic_{r,2}^z)$.
As our representation of spin operator implies~\cite {Baskaran07}, each $s^z$ inserts a $Z_2$ flux into neighbouring plaquetes. 
Hence, average in Eq. (\ref{h4}) does not vanish only if spin operators in this expression come in pairs, so that 
zero-flux state is obtained after all spin operators have been acting upon the ground state. 
These pairs of spin operators come in general at different time moments.  However,
for the time interval when such a flux exists in the intermediate state, a potential of the order of $J$ for Majorana fermions is turned on~\cite {TFK2011}. Contribution of such intermediate states is thus suppressed.
This way, time indices become paired, too. In order to illustrate this mechanism, consider a contribution to the expression
 (\ref{h4}) for $r=r_1$, $r_2=0$ and $r_3=r_4=\rho$.
In terms of the bond fermions ($\psi_r^\alpha=(\psi,\psi^+)$,$\phi_r^\alpha=(\phi^+,\phi)$, this expression (for $t_1<0$) reads (analogously to ~\cite {TFK2011}) :
\begin{gather}
\label{example2}
\langle T s^z_r(t)s^z_0(0)s^z_{r_1}(\tau_1)\ldots s^z_{r_4}(\tau_4)\rangle = \\ \nonumber
2^5\langle  e^{i \hat H t}\psi^{\alpha_1}_r\phi^{\alpha_1}_r e^{-i \hat H t} \psi^{\alpha_2}_0\phi^{\alpha_2}_0 e^{i \hat H \tau_1}\psi^{\alpha_3}_r\phi^{\alpha_3}_r e^{-i \hat H \tau_1}e^{i \hat H \tau_2}\psi^{\alpha_4}_0\phi^{\alpha_4}_0 e^{-i \hat H \tau_2}e^{i \hat H \tau_3}\psi^{\alpha_5}_\rho\phi^{\alpha_5}_\rho e^{-i \hat H \tau_3}e^{i \hat H \tau_4}\psi^{\alpha_6}_\rho\phi^{\alpha_6}_\rho e^{-i \hat H t}\rangle=\\ \nonumber
-2^5\langle  e^{i \hat H t}\psi_r e^{-i \hat H_r t} \psi_0 e^{i \hat H_{r,0} \tau_1} \psi^+_r  e^{-i \hat H_0 \tau_1} e^{i \hat H_0 \tau_2}\psi^+_0 e^{-i \hat H \tau_2}e^{i \hat H \tau_3}\psi_\rho e^{-i \hat H_\rho \tau_3}e^{i \hat H_\rho \tau_4}\psi_\rho e^{-i \hat H t}\rangle=\\ \nonumber
2^5\langle T \psi_r(t)\psi^+_r(\tau_1) \psi_0(0)\psi^+_0(\tau_2)\psi_\rho(\tau_3)\psi_\rho(\tau_4)e^{-i\int \hat V(\tau)d \tau} \rangle
\end{gather}
with
\begin{equation}
V(\tau)=\theta(\tau-\tau_1)\theta(t-\tau)\hat V_r+\theta(\tau-\tau_2)\theta(0-\tau)\hat V_0+\theta(\tau-\tau_3)\theta(\tau_4-\tau)\hat V_\rho.
\end{equation}
where $V_r=4J(\psi^+_r\psi_r-\frac{1}{2})$ and $H_r=H+V_r$.
In the series of transformations shown in Eq (\ref{example2}), we have used:
 i)  the relation between spin and fermionic operators was used for the transformation from line $1$ to line $2$; 
ii) commutation relations  $\phi e^{iHt}=e^{iH_rt}\phi$ and $\phi^+ e^{iH_rt}=e^{iHt}\phi^+$ was used to transform
line $2$ into line $3$,
 and iii) identities  $\phi^+\phi|0>=\frac{1+u_{ij}}{2}|0>$ and $e^{iHt}e^{-iH_rt}=Te^{-i\int V(\tau)d\tau}$
have been employed to obtain finally line $4$. 

Following the standard route\cite{FES}, we factorize the expression (\ref{example2}) as $2^5 e^{C}L$, where $C$ stays for the sum of connected diagrams, $e^C=\langle T e^{-i\int \hat V(\tau)d \tau} \rangle$ and $L$ stays for the contribution
of 'connected line'. 
In the case of $t-\tau_1$ of the order of $t$ this expression oscillates
at high frequency $\sim J$, suppressing the value of the integral over $\tau_1$. As a result, the dominating contribution is expected to come from a region of $t\approx \tau_1$, but if $t>0>\tau_1$, such a region is absent
and the whole contribution in Eq.(\ref{example2})  is small. 
On the contrary, such a suppression does not occur for the region  $t>\tau_1>0$.  Here integration over $\tau_1$ 
is  equivalent to  the substitution $\tau_1\to t$ and  multiplying the result by a additional factor 
$(i h_0)^{-1} $, where $h_0 \sim J$, see~\cite {TFK2011} for details.

 All other pairings of spin operators can be considered similarly. As a result, the whole contribution 
to spin-spin correlation function of the order $h_z^4$
can be represented in the form of (the integral of)  three-point correlation function of fermionic density:
\begin{equation}
\left\langle s_{r}^{z}(t)s_{0}^{z}(0)\right\rangle^{(4)} =
2^6\cdot 4\cdot 
\frac{h_z^4}{(i h_0)^3}\sum\limits_{\rho}\int d\tau \langle T \hat n^T(t,r) \hat n^T(0,0)\hat n^T(\tau,\rho)\rangle
\label{s1}
\end{equation}
where $ \hat n^T(t,r)=\hat \psi(t,r)\hat \psi^+(t,r)$. 
The factor of $4$ result from two permutations of $s^z_t(r)$ and the paired spin operator giving the same contribution 
(the same for $s^z_0(0)$).  
Below we introduce short-hand notation $ x = (\mathbf{r}, t)$.
\begin{figure}[tbp]
\includegraphics[width=12cm,height=4cm]{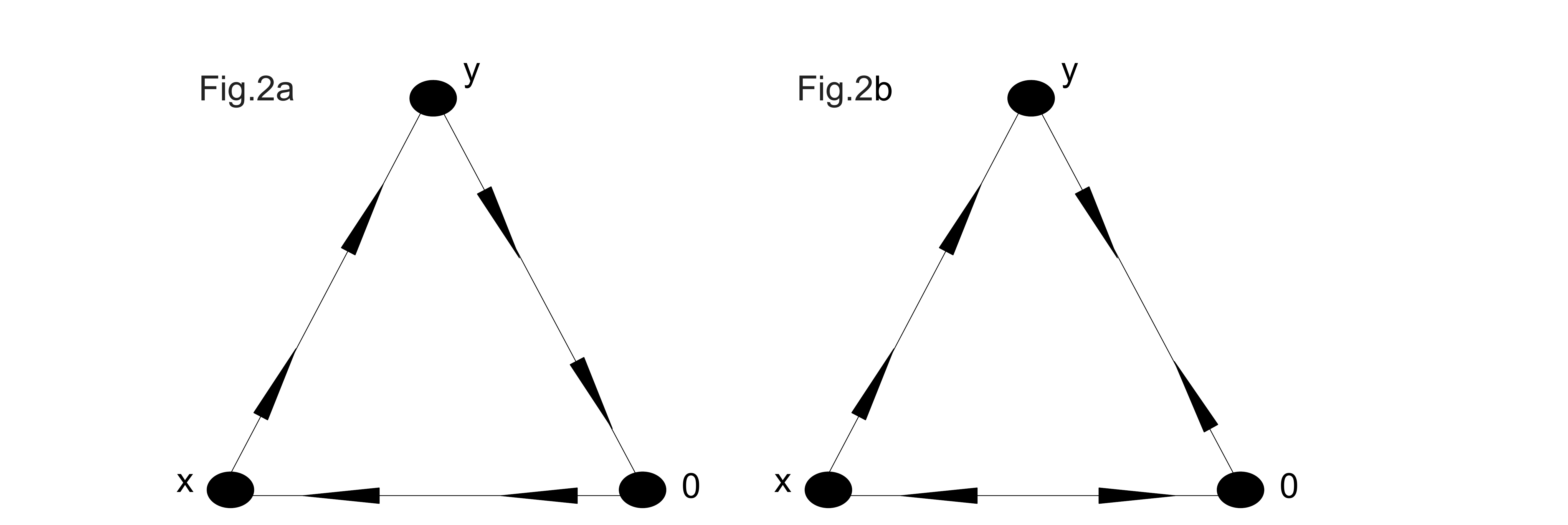}
\caption{Two types of diagrams, contributing to the fourth-order correction to the spin-spin correlation function.}
\label{pdiag}
\end{figure}

\emph{Evaluation of the Fermionic correlation function.}
Writing $x = (r, t)$, it is convenient to define two Green functions\cite{TFK2011}: $G(x)=\langle T\psi(x)\psi^+(0)\rangle=-\langle T\psi^+(0)\psi(x)\rangle$ and $F(x)=\langle T\psi(x)\psi(0)\rangle=\langle T\psi^+(0)\psi^+(x)\rangle$. Explicitely (for $t>0$):
\begin{equation}
G(x)=\frac{-\sqrt{3}Jt (Z_1+Z_2)+(Z_2-Z_1)r\cos\gamma}{4\pi(3J^2t^2-r^2)^{3/2}}
\end{equation}
and
\begin{equation}
F(x)=-i\frac{(Z_1+Z_2)r\sin\gamma}{4\pi(3J^2t^2-r^2)^{3/2}}.
\end{equation}
With these equations, the expectation value in expression (\ref{s1}) can be evaluated with the use of Wick theorem. Evaluation of the diagrams
 (see few examples in Fig \ref{pdiag}) requires calculation of convolutions of pairs of Green functions. As a result, combining all the contributions, we get:
for the density-density correlation function contribution of the order $h_z^4$: 
\begin{equation}
\label{fin}
S^{(4)}(\mathbf{r},t)= \frac{h_z^4}{2 (ih_0)^3} \frac{2^6}{\pi^2\sqrt{3}}
\frac{ r\cos\gamma(Z_2^2-Z_1^2)}{J(3J^2t^2-r^2)^2} = 
\frac{h_z^4}{h_0^3} \frac{2^6}{\pi^2\sqrt{3}}
\frac{ r\cos\gamma\sin(\frac{2}{3}m\pi)}{J(3J^2t^2-r^2)^2}.
\end{equation}
Note that the ratio of the  new contribution (\ref{fin}) to spin-spin correlation function
to the lowest-order one given by Eq.(\ref{res2}),  grows $\propto (h_z/J)^2  r$ at large distances.
Therefore the result (\ref{fin}) is applicable at $r \ll (J/h_z)^2$.

\emph{Generalization to higher orders in magnetic field.}

Perturbative contribution (\ref{fin}) can be reproduced by the lowest-order correction over the weak 
vector potential
$\mathbf{A}$   minimally coupled to the gradient term in the low-energy Fermionic Hamiltonian (\ref{Dirac}):
\begin{equation}
H_F = \sqrt{3} J \sum_{\mathbf{p}} \bar{a}_\lambda(\mathbf{p})\sigma^\alpha_{\lambda\mu} (p^\alpha - A^\alpha)
  a_\mu(\mathbf{p})  \quad {\rm where} \quad \mathbf{A} = (-\delta, 0) \quad {\rm and} \quad \delta=\frac{4 h_z^2}{\sqrt{3}h_0 J}
\label{DiracA}
\end{equation}
Below we show that expression (\ref{DiracA}) is in fact more general: it accounts for all orders of expansion in powers of $(h_z/J)^2$.

Consider $2n$-th order of perturbation theory over $h_z$ for the spin-spin correlation function:
\begin{equation}
\langle s^z_{r}(t)s_0^z(0)\rangle^{(2n)}=\frac{(-i h_z)^{2n}}{2n!}\sum\limits_{r_1,\ldots,r_{2n}}\int d \tau_1\ldots d \tau_{2n} \langle T s^z_r(t)s^z_{r_1}(\tau_1)\ldots s^z_{r_{2n}}(\tau_{2n})s^z_{0}(0) \rangle
\label{n}
\end{equation}  
In order for the  correlation function under the integral to be  nonzero,
$2n$  spin operators  in this expression should  come in $n$ pairs, like $s^z(r_j,t_j) s^z(r_j, t_j')$ .  This is necessary in order
that all fluxes created by the action of these spin operators  will be  eliminated eventually \cite{Baskaran07}.   Thus summation in Eq.(\ref{n})
goes over $n$ independent sites $r_i$.

We represent spin operators using the relation $s^z_r=2i\psi_r^\alpha\phi_r^\alpha$. Using commutation relations and  the identity $\phi^+\phi|0\rangle=\frac{1+u_{ij}}{2}|0\rangle$  we transform the original correlation function into the form,  which includes only $\psi$-operators,
 like in  the last line of Eq.(\ref{example2}) \cite{TFK2011}. The cost of this transformation is that now we deal with
a correlation function defined for a more involved  Hamiltonian  including the  action of time-dependent fluxes.
 Fluxes are created by the action  one spin  operator from any pair and destroyed after some time $\tau$ by the action of
another spin operator.  Typical values of  $\tau$  are smaller then the time periods between two appearances of a flux,  so  we 
can integrate over corresponding time-off (and time-on)  of each  of these fluxes separately.  
It brings us to the expression similar to (\ref{s1}):
\begin{eqnarray}
\langle s^z_{r}(t)s_0^z(0)\rangle^{(2n)}=16(-1)^{n-1}\frac{(-i 2h_z)^{2n}}{(i h_0)^{n+1}n!}\sum_{r_1,r_2...r_{n-1}}\int d\tau_1...d\tau_{n-1}\langle T  \psi_{r_1}(\tau_1)\psi^\dagger_{r_1}(\tau_1)...\nonumber \\ \psi_{r_{n-1}}(\tau_{n-1})\psi^\dagger_{r_{n-1}}(\tau_{n-1})\psi_{r}(t)\psi^\dagger_{r}(t)\psi_{0}(0)\psi^\dagger_{0}(0)\rangle
\nonumber \\
\label{n2}
\end{eqnarray}

Consider now  perturbation series for the Fermionic density correlation function
\begin{eqnarray}
D(\mathbf{r},t) = \langle T\psi_{r}(t)\psi^\dagger_{r}(t)\psi_{0}(0)\psi^\dagger_{0}(0) \rangle 
\label{Drt}
\end{eqnarray}
defined with the Hamiltonian $H_{new}=H_F + \mathcal{V}$, where  $H_F$ is defined in Eq.(\ref{Dirac}) and  
$\mathcal{V}=\frac{4 h_z^2}{h_0}\sum\limits_{r} \psi_r\psi^\dagger_r$.  It is easy to see  that 
expansion of  (\ref{Drt}) over $\mathcal{V}$ produces (up to the overall coefficient  $\frac{64h_z^2}{h_0^2}$) 
 the same perturbative  series as the one presented in Eq.(\ref{n2}). Therefore we conclude that
\begin{eqnarray}
\langle s^z_{r}(t)s_0^z(0)\rangle=\frac{64h_z^2}{h_0^2}\langle T\psi_{r}(t)\psi^\dagger_{r}(t)\psi_{0}(0)\psi^\dagger_{0}(0) \rangle_{new}
\end{eqnarray}
The Hamiltonian $H_{new}$ is still quadratic in Fermions and admits two equivalent interpretations: a) perturbation $\mathcal{V}$ can be
rewritten in terms of vector potential, exactly like in Eq.(\ref{DiracA}), and b) one can understand the effect of $\mathcal{V}$
as the shift of the conic points of Fermionic spectrum according to:
 $\mathbf{K}^{*}_{1,2}=\mathbf{K}_{1,2}\mp (\delta,0) $ where $\mathbf{K}_{1,2}$ are conic points 
of $H_F$ and $\delta$ was determined in Eq. (\ref{DiracA}).  
Therefore final result for the dynamic spin-spin correlation function $S(\mathbf{r},t)$ is provided by Eq.(\ref{res2})
where $Z_{1,2}$ are replaced by $Z^*_{1,2} = e^{i \mathbf{K}^*_{1,2} \mathbf{r}}$.

\emph{Dynamical structure factor.}

The dynamical spin structure factor, $S(\mathbf{p},\omega)$, is determined by the Fourier transform of the spin correlation function 
$S(\mathbf{r},t)=\langle s^z_{r}(t)s_0^z(0)\rangle$ and can be measured by inelastic neutron scattering. 
We have calculated  $S(\mathbf{r},t)$ in the asymptotic region  $r\gg1$, $t\gg\frac{1}{J}$ and $|r-\sqrt{3}Jt|\gg 1$. 
It allows to calculate the singular dependencies of  $S(\mathbf{p},\omega)$ on low frequency $\omega$ and momentum $\mathbf{p}$ located near $\mathbf{p}=0$ and 
$\mathbf{p}_K=\mathbf{K}_1-\mathbf{K}_2=(\frac{4\pi}{3}-2\delta,0)$.

 At small $\mathbf{p}$ we obtain:
 \begin{eqnarray}
\label{S1}
 S(\omega,\mathbf{p})-S(0,0)=\frac{4h^2_z}{3J^2h_0^2}\left(\sqrt{\omega^2-3J^2p^2}+\frac{3J^2p^2\cos^2(\phi)}{\sqrt{\omega^2-3J^2p^2}}\right)
 \end{eqnarray}
Here  $\phi$ is polar angle between $\mathbf{p}$ and $\mathbf{x}$. 
This expression is applicable for small $w$ and $p =|\mathbf{p}|$ and the condition  $w^2 > 3J^2p^2$.
At $w^2 < 3J^2p^2$ the same approximation leads to zero result. 

Near $\mathbf{p}_{K}$ we find:
 \begin{eqnarray}
 \label{S2}
 S(\omega,\mathbf{p})-S(0,\mathbf{p}_{K})=\frac{4h^2_z}{3J^2h_0^2}\sqrt{w^2-3J^2(\mathbf{p}-\mathbf{p}_{K})^2}
 \end{eqnarray} 
 where $w^2 > 3J^2p^2$.

Our results (\ref{S1},\ref{S2}) for the structure factor differ considerably from those obtained in Ref.~\cite{song2016low} 
for another version of Kitaev model with local perturbations. In our case, the structure factor is  proportional $\omega$ when $\mathbf{p}\to 0,\mathbf{p}_{K}$ unlike $\omega^3$ obtained in Ref.~\cite{song2016low}. This difference is connected with the lacking time reversal symmetry in our model: starting from spin-spin correlation function we get density-density correlator which does not vanish near the conical points.

Note that in this Letter, we have assumed the magnetic field to be directed along $z$ axis. Nevertheless, our treatment can be generalized to arbitrary direction of the magnetic field
 as long as it has vanishing mixed product $h_xh_yh_z$ and low-energy Majorana Hamiltonian is massless\cite{Kitaev06}. In this case, the replacement of low-energy effective Hamiltonian (\ref{Dirac}) by its extended version (\ref{DiracA}) with appropriate vector potential $\mathbf{A}$ captures all effects of nonzero magnetic
field. In the case of $h_xh_yh_z\ne 0$, correlation function $S(r,t)$ is expected to decrease exponentially at $r \gtrsim \frac{1}{h_xh_yh_z}$.

We are grateful to A. Yu. Kitaev for very useful discussions. This research was supported by the Russian Science Foundation
 grant \# 14-12-00898.

\vspace{-0.5cm}

\bibliographystyle{apsrev}
\bibliography{qsl.bib}

\end{document}